\documentclass[twocolumn,preprintnumbers,amsmath,amssymb]{revtex4}

\usepackage{graphicx}
\usepackage{dcolumn}
\usepackage{bm}
\usepackage{epstopdf}

\begin{document}


\title{Terahertz oscilloscope for recording time information of ultrashort electron beams}

\author{Lingrong Zhao$^{1,2}$, Zhe Wang$^{1,2}$, Heng Tang$^{1,2}$, Rui Wang$^{1,2}$, Yun Cheng$^{1,2}$, Chao Lu$^{1,2}$, Tao Jiang$^{1,2}$, Pengfei Zhu$^{1,2}$, Long Hu$^{3}$, Wei Song$^{3}$, Huida Wang$^{3}$, Jiaqi Qiu$^{4}$, Roman Kostin$^{5}$, Chunguang Jing$^{5}$, Sergey Antipov$^{5}$, Peng Wang$^{6}$, Jia Qi$^{6}$, Ya Cheng$^{6,7}$, Dao Xiang$^{1,2,8*}$ and Jie Zhang$^{1,2*}$}
\affiliation{%
$^1$ Key Laboratory for Laser Plasmas (Ministry of Education), School of Physics and Astronomy, Shanghai Jiao Tong University, Shanghai 200240, China \\
$^2$ Collaborative Innovation Center of IFSA (CICIFSA), Shanghai Jiao Tong University, Shanghai 200240, China \\
$^3$ Science and Technology on High Power Microwave Laboratory, Northwest Institute of Nuclear Technology, Xi'an, Shanxi 710024, China\\
$^4$ Nuctech Company Limited, Beijing, 100084, China\\
$^5$ Euclid Techlabs LLC, Bolingbrook, Illinois 60440, USA \\
$^6$ State Key Laboratory of High Field Laser Physics, Shanghai Institute of Optics and Fine Mechanics, Chinese Academy of Sciences, Shanghai 201800, China\\
$^7$ State Key Laboratory of Precision Spectroscopy, East China Normal University, Shanghai 200062, China\\
$^8$ Tsung-Dao Lee Institute, Shanghai 200240, China\\
}
\date{\today}

\begin{abstract} 
We propose and demonstrate a Terahertz (THz) oscilloscope for recording time information of an ultrashort electron beam. By injecting a laser-driven THz pulse with circular polarization into a dielectric tube, the electron beam is swept helically such that the time information is uniformly encoded into the angular distribution that allows one to characterize both the temporal profile and timing jitter of an electron beam. The dynamic range of the measurement in such a configuration is significantly increased compared to deflection with a linearly polarized THz pulse. With this THz oscilloscope, nearly 50-fold longitudinal compression of a relativistic electron beam to about 15 fs (rms) is directly visualized with its arrival time determined with 3 fs accuracy. This technique bridges the gap between streaking of photoelectrons with optical lasers and deflection of relativistic electron beams with radio-frequency deflectors, and should have wide applications in many ultrashort electron beam based facilities. 
\end{abstract}

\maketitle

The ability to characterize the time information of an ultrashort electron beam including both the temporal profile and arrival time is crucial for optimizing and enhancing the performance of many electron beam based scientific facilities such as free-electron lasers (FELs \cite{LCLS, SACLA, PAL}), ultrafast electron diffraction (UED \cite{UED1, UED2}) and microscopy (UEM \cite{UEM1, UEM2, UEM3, UEM4}), laser-driven and beam-driven advanced accelerators \cite{LPARMP, PWFANature, DLARMP, THzNC, THzNP, THzIFEL}, etc. In accelerator community, radio-frequency (rf) deflecting cavities have been widely used to measure the temporal profile of relativistic electron beams with energy ranging from MeV to GeV (see, e.g. \cite{TCAVDESY, TCAVSLAC, CUCLA}). However, the information of beam arrival time with respect to an external laser as required in a pump-probe experiment can not be directly measured with an rf deflector. In attosecond science community, streaking of photoelectrons with optical lasers has become a standard technique for characterizing the complete information of attosecond pulses \cite{AS1}. Recently, this technique has been adapted to characterize femtosecond x-ray pulses in FELs with the streaking imprinted by far-infrared and THz pulses \cite{TS1, TS2, TS3, TS4}. However, this technique doesn't apply to a relativistic electron beam, as dictated by Lawson-Woodward theorem \cite{PRE}. Very recently, THz streaking of keV and MeV electrons \cite{THzbuncher, SJTUstreaking, SLACstreaking} in a sub-wavelength metallic structure has been used to measure both the temporal profile and arrival time of electron beams. However, the small aperture used to enhance THz field may significantly limit the number of useful electrons. Furthermore, with the streaking imprinted by a linearly polarized THz pulse, the beam receives sinusoidal angular streaking and thus the measurement has a rather limited dynamic range (time window where the measurement is accurate) comparable to about one quarter of the wavelength.

    \begin{figure*}[t]
    \includegraphics[width = 0.8\textwidth]{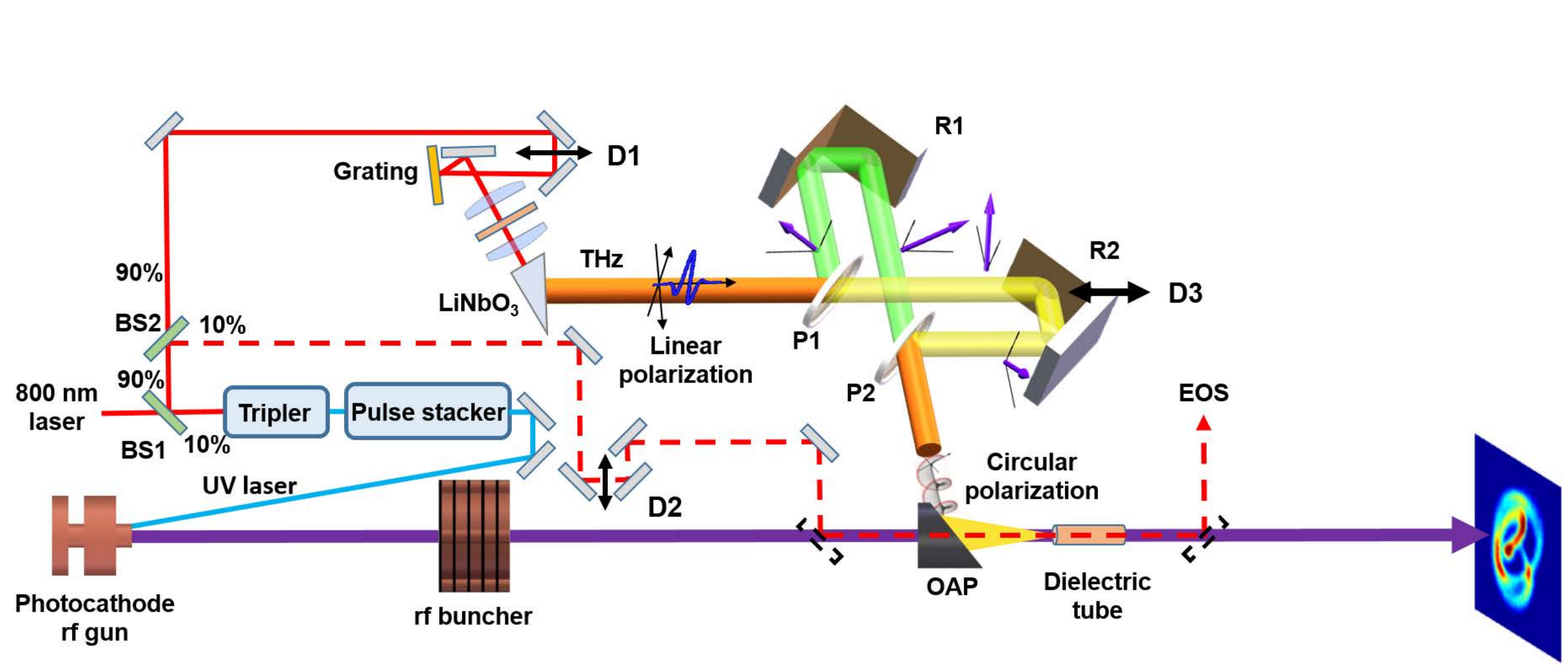}
            \caption{THz oscilloscope experiment setup. The electron beam is produced in a photocathode rf gun with a UV laser. A pulse stacker consisting of a set of BBO crystals is used to shape the laser temporal profile. The electron beam can be longitudinally compressed with the rf buncher and the time information of the electron bunch can be inferred from the THz-induced deflection in the dielectric tube. The polarization of the THz pulse is manipulated with polarizers (P1 and P2) and roof mirrors (R1 and R2). The THz pulse is focused to the dielectric tube with an off-axis parabolic mirror (OAP).  
    \label{Fig.1}}
    \end{figure*}

In this Letter, we demonstrate a laser-driven THz oscilloscope that allows one to record the complete time information of an ultrashort electron beam with both large dynamic range and high temporal resolution. This scheme exploits the sustained interaction between the THz pulse and the electron beam in a dielectric tube, allowing sufficient defletcion to be generated without enhancing the THz field with a narrow slit. With accurate control of the THz polarization, the electron beam is deflected helically, mapping the time axis uniformly into angular axis. With this THz oscilloscope, the process of nearly 50-fold longitudinal compression of a relativistic electron beam to about 15 fs (rms) is visualized with its arrival time determined with 3 fs (rms) accuracy. Such THz oscilloscope should have wide applications in many electron beam based facilities.

The setup of THz oscilloscope experiment is shown in Fig.~1. The electron beam is produced in a 1.5 cell S-band (2856 MHz) photocathode rf gun. The beam energy is about $3.4$ MeV at the gun exit and it may be longitudinally compressed by a C-band (5712 MHz) rf buncher \cite{SJTUstreaking}. The THz pulse is produced with a $\sim$5 mJ laser in LiNbO$_3$ crystal with the tilted-pulse-front-pumping scheme \cite{TPFP} and further injected into a dielectric tube where it interacts with the electron beam and changes the transverse momentum of the electrons. Two wire grid polarizers and roof mirrors are used to manipulate THz polarization to control the deflection pattern of the beam. The THz-induced deflection is measured at a screen 1.7 m downstream of the dielectric tube. 

The maximal THz energy just after the crystal is measured to be about 3 $\mu$J with a calibrated Golay cell detector. The THz transverse size at the entrance to the dielectric tube is measured to be about 1 mm (rms) with a THz camera. The THz waveform measured with electro-optic sampling (EOS) technique \cite{EOS1} at the entrance to the dielectric tube is shown in Fig.~2a. In contrast to a plane wave which only interacts with a relativistic electron beam very weakly because the electric force largely cancels the magnetic force, the field excited by the THz pulse in a dielectric tube can support effective interaction between the THz pulse and relativistic electron beam \cite{DLARMP}. In our experiment, a 5 mm long cylindrical quartz tube with inner (outer) diameter of 870 (930) $\mu$m is used. The outer surface of the tube is gold coated and analysis with CST microwave studio software shows that it can support HEM11 mode with frequency at about 0.6 THz suitable for streaking beam transversely. The field distribution and the way the beam is deflected depends primarily on the polarization of the injected THz pulse.

    \begin{figure}[b]
    \includegraphics[width = 0.49\textwidth]{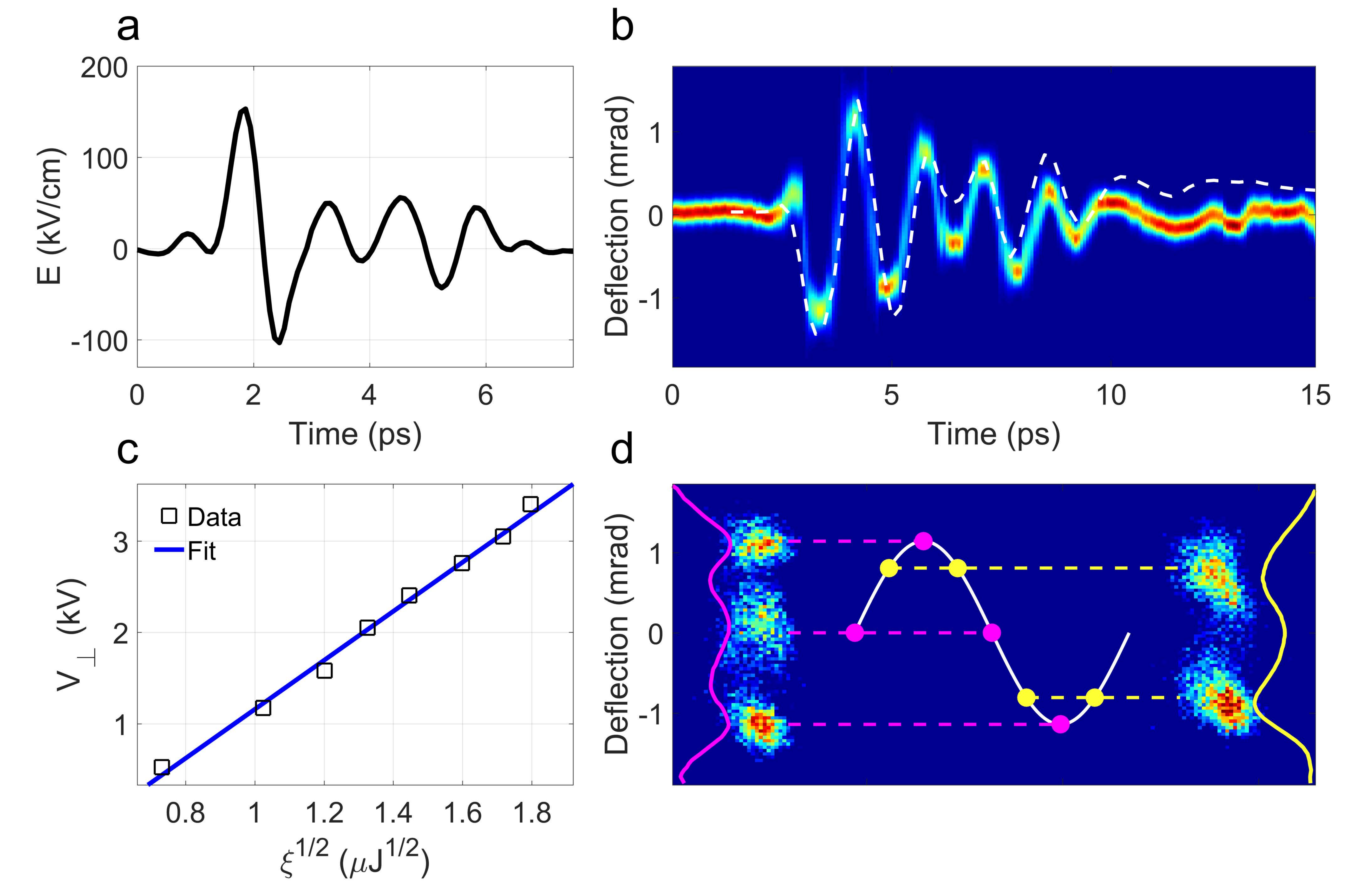}
            \caption{Measured THz waveform with EOS (a); Measured and simulated (white dashed line) beam deflection as a function of time delay between the electron beam and linearly polarized THz pulse (b); Measured maximal deflection vs THz energy (c); Deflected beam distribution for a train of four equally spaced bunches (d). In (d) the measurement is done at two different conditions where the bunch trains (magenta and yellow dots) ride at different phases of the THz field.    
    \label{Fig.2}}
    \end{figure}

      \begin{figure*}[t]
    \includegraphics[width = 0.89\textwidth]{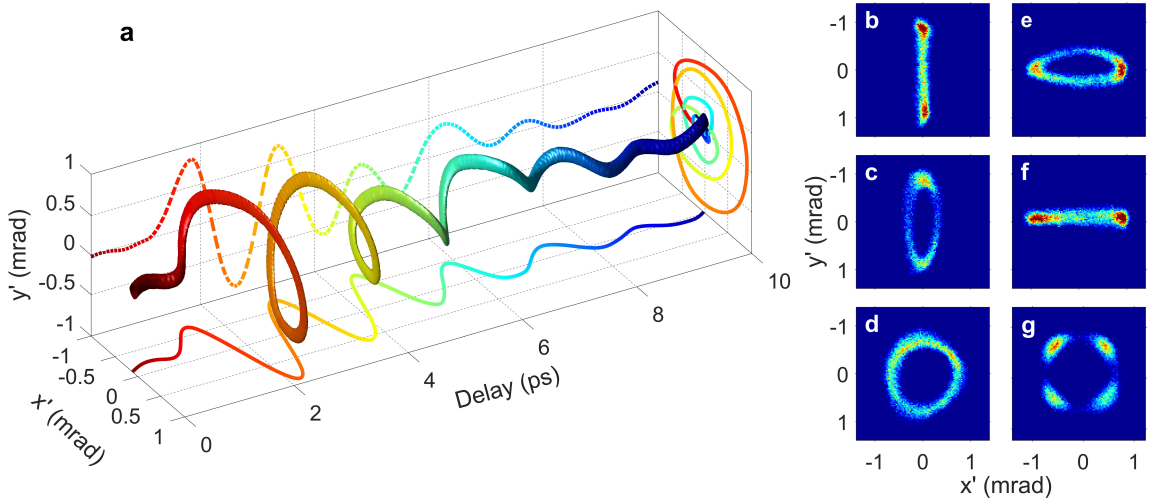}
            \caption{(a) Measured electron beam 3D deflectogram with circularly polarized THz pulse. Measured deflected beam distribution when the relative phase of the two THz pulses in the Martin-Puplett interferometer is set at $\phi=0$ (b), $\phi=\pi/4$ (c), $\phi=\pi/2$ (d), $\phi=3\pi/4$ (e) and $\phi=\pi$ (f); Deflected beam distribution for a train of four equally spaced bunches (g).
    \label{Fig.3}}
    \end{figure*}   

After both the spatial and temporal overlap between the THz and the electron beam is achieved in the dielectric tube \cite{SJTUstreaking}, the timing of the THz beam is varied (in 67 fs steps) and the measured beam deflectogram (Fig.~2b) is found to be in reasonable agreement with simulation (with CST particle-in-cell solver). With the group velocity of the THz pulse being about 0.8$c$, the interaction window (a time window within which the THz pulse can effectively interact with the electron beam) is larger than the THz pulse width (see, e.g. \cite{Jamison}). In this measurement, the electron pulse width is about 150 fs (rms) and each measurement is obtained with integration over 10 pulses. The maximal deflection voltage $V_\perp$ calculated from the measured deflection angle ($y'=V_\perp/E$, where $E$ is the electron beam energy) for various THz energies are shown in Fig.~2c where one can see that $V_\perp$ is proportional to the square root of the THz energy, implying that the interaction is a linear process. 

This time-dependent deflection allows one to map the electron beam time information into angular information which is further converted into spatial distribution on a downstream screen. Therefore, both the electron beam temporal profile and arrival time can be measured. However, the dynamic range is limited to about one quarter of the wavelength where the deflection strength is approximately linear. To illustrate this effect, we used two BBO crystals to shape the UV pulse for producing a train of four bunches equally separated by about 0.4 ps \cite{CS}. Because the electron bunch train is longer than the dynamic range, the measured distribution of the bunch train (Fig.~2d) shows considerable distortions which lead to ambiguities in retrieval of the bunch temporal structure. 

To increase the dynamic range and to make the streaking strength independent of the phase, a circularly polarized pulse may be used \cite{ZZ, TS1}. However, the conventional quarter-wave plate that converts linearly polarized radiation into circularly polarized pulse only works for narrow-band pulses. Here, we use a pair of wire grid polarizers and roof mirrors for such a conversion (total transmission efficiency $>$95\%), similar to the configuration of a Martin-Puplett interferometer (see, e.g. \cite{MP}). As shown in Fig.~1, the vertically polarized THz pulse first enters a polarizer (P1) with wires oriented at 45 degrees relative to the beam axis and is split into two pulses with orthogonal polarization. The polarization of each THz pulse is then rotated by 90 degrees with the roof mirror and finally the two THz pulses are recombined at the second polarizer (P2) before injection into the dielectric tube. With the phase difference of the two pulses set to $\pi/2$, the measured 3D beam deflectogram is shown in Fig.~3a. The 3D deflectogram may be understood as superposition of two orthogonal deflections, and the deflectograms projected to $x-z$ and $y-z$ planes (solid and dashed curves; only the beam centroid is shown) also clearly show $\pi/2$ phase difference. The deflectogram projected to $x-y$ plane represents integration over 10 ps time window and thus shows spiral pattern (the time information is represented by the color of the spiral line). 

It should be noted that the relative phase of the two THz pulses can be continuously varied by changing the path length. This provides an effective way to switch the superimposed THz pulse polarization, allowing various deflection patterns to be formed, similar to a cathode-ray oscilloscope. For instance, in a separate experiment we produced a flat-top beam with about 1.6 ps pulse width matching the wavelength of the deflection field, and the measured deflection patterns for various phase differences of the two pulses are shown in Fig.~3b-f. Specifically, linear deflections are achieved when the phase delay is at 0 (Fig.~3b) and $\pi$ (Fig.~3f). Circular deflection is achieved when $\phi=\pi/2$ (Fig.~3d) and elliptical deflections are achieved at other phases (Fig.~3c for $\phi=\pi/4$ and Fig.~3e for $\phi=3\pi/4$). The circular deflection also increases the dynamic range and maps the time information more uniformly into the angular distribution. For the same bunch train as used in Fig.~2d, the measured beam distribution with the circular deflection mode is shown in Fig.~3g where one can see that the four bunches are indeed uniformly distributed along the angular axis.   

      \begin{figure}[b]
    \includegraphics[width = 0.49\textwidth]{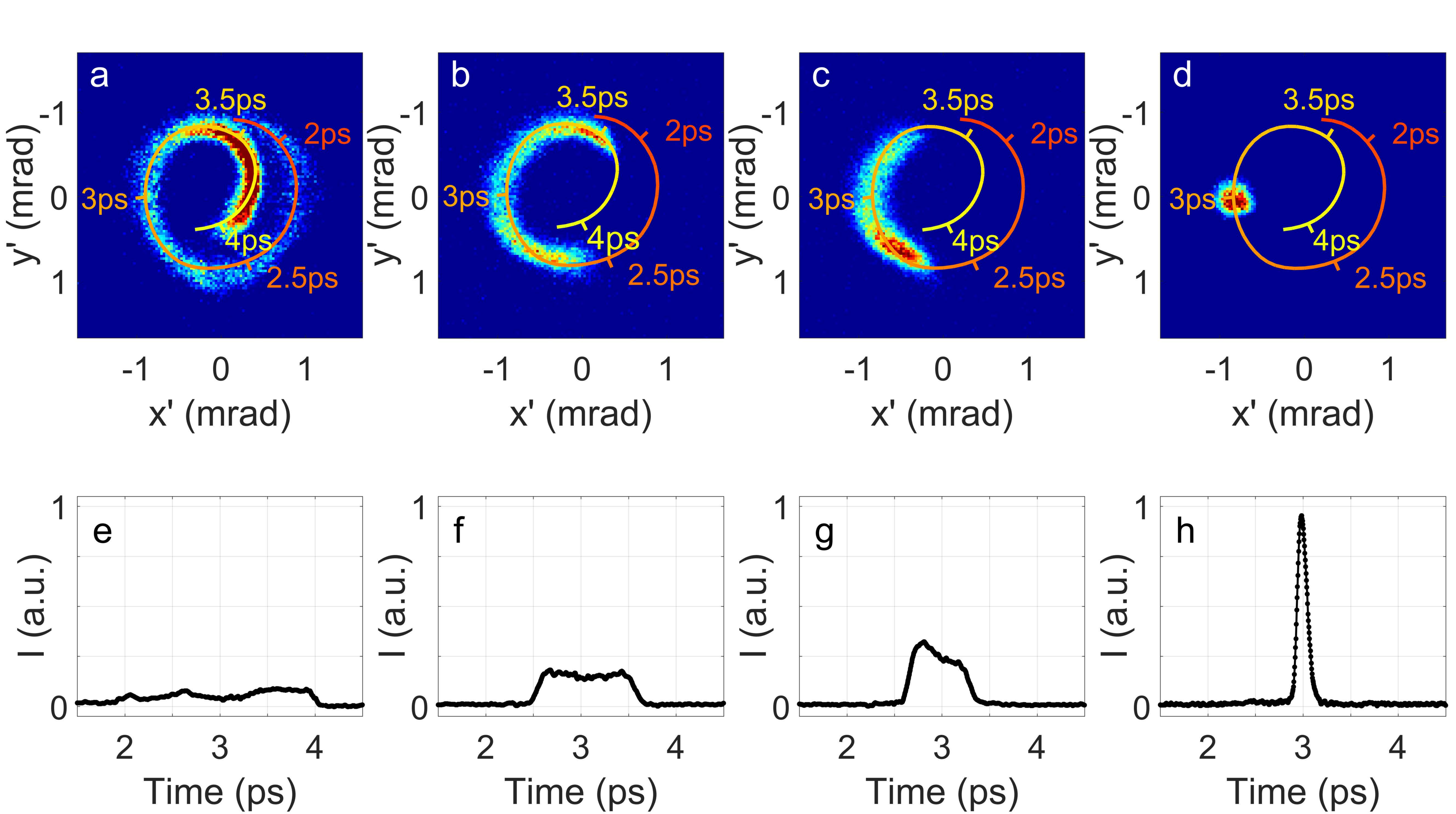}
            \caption{Streaked electron beam distribution measured with THz oscilloscope when the buncher voltage is set to $V_b=0$ (a), $V_b=0.5~$MV (b), $V_b=0.8~$MV (c) and $V_b=1~$MV (d). The corresponding beam temporal profiles after converting the angular distribution into time are shown in the bottom row. The number of electrons in the bunch is about $3\times10^5$.  
    \label{Fig.4}}
    \end{figure}   

It should be noted that taking advantage of the ramping of the deflection envelope where the spiral lines are well separated in certain regions (e.g. from 1.5 ps to 4.5 ps in Fig.~3a), the dynamic range of the measurement can be further increased beyond the wavelength of the deflection field \cite{TS4, ZZ}. For instance, Fig.~4a shows the deflected beam distribution when the beam full width is about 50\% longer than the wavelength of the deflection field where a spiral pattern is formed. Such an increased dynamic range allows us to directly visualize the process of nearly 50-fold longitudinal bunch compression. In this measurement, the electron beam passes through the C-band buncher cavity where the bunch head is decelerated and the bunch tail is accelerated. Further sending the beam through a drift allows the bunch tail to catch up with the bunch head, and hence longitudinal bunch compression is achieved \cite{CPRL, CUCLA, SJTUstreaking}.

The temporal resolution in measuring the beam temporal profile with this THz oscilloscope may be approximately estimated as $\delta t=\sigma' T/2\pi \theta$, where $\sigma'$ is the intrinsic beam divergence, $T$ is the period of the defection, $\theta$ is the maximal deflection angle. In our experiment, the transverse beam size at screen P1 was measured to be about 170 $\mu$m with the THz off, corresponding to $\sigma'=100~\mu$rad and the temporal resolution is estimated to be about 24 fs. Fig.~4b-d shows deflected beam profiles for increasing voltages of the buncher where one can clearly see the shrinking of the beam profile towards the bunch center. The corresponding electron beam temporal profiles are shown in Fig.~4e-h. The raw beam pulse width at full compression is measured to be about 28 fs (rms), limited by the resolution of the measurement. Subtracting the resolution in quadrature yields a true bunch length of about 15 fs (rms), indicating nearly 50-fold longitudinal compression (initial bunch length is about 700 fs (rms)). 

      \begin{figure}[b]
    \includegraphics[width = 0.49\textwidth]{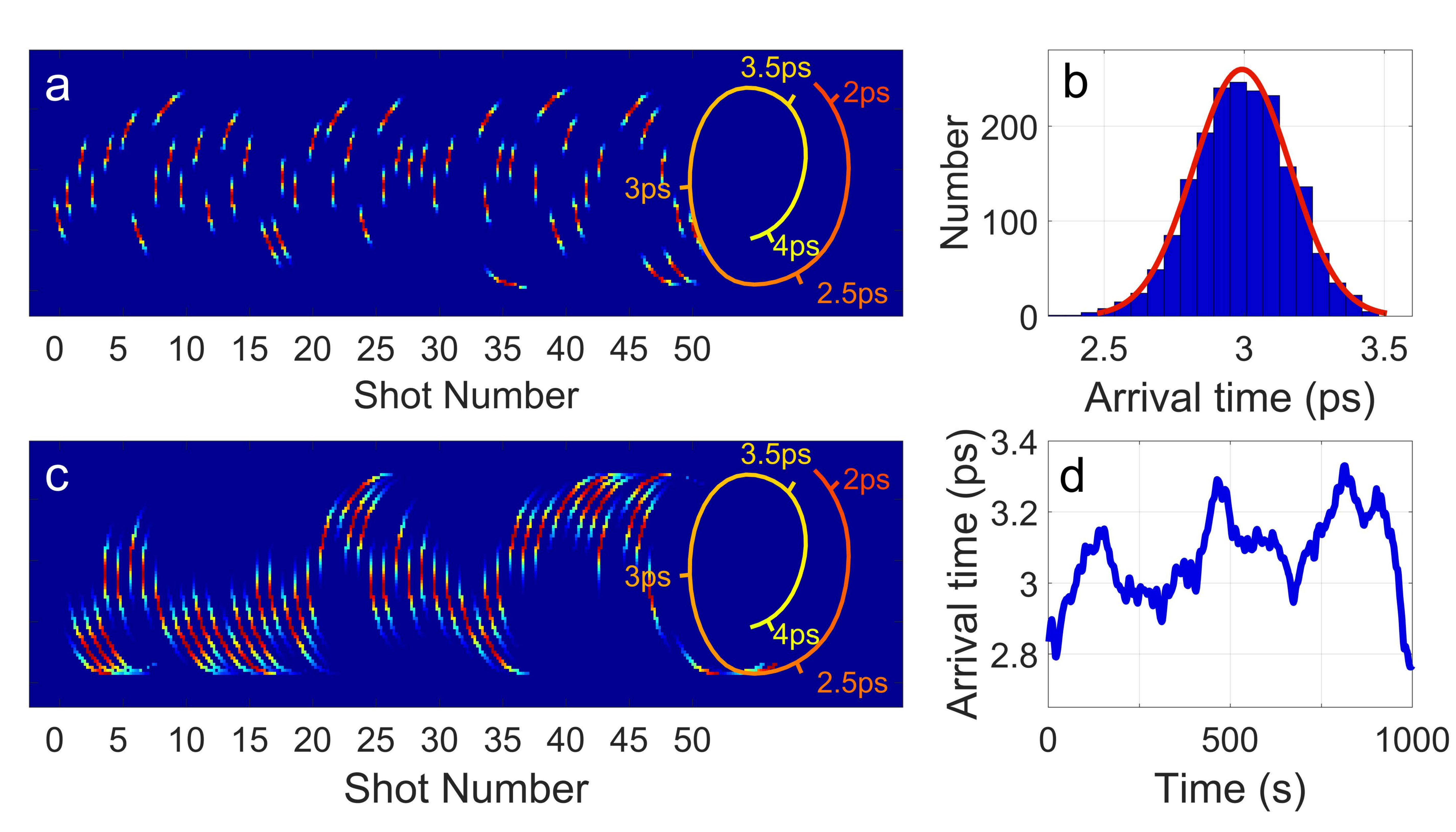}
            \caption{(a) 50 consecutive measurement of beam arrival time using the THz oscilloscope (time scales shown with solid curve); (b) Distribution of the beam arrival time collected over 2000 shots. A Gaussian fit (red line) to the distribution yields a jitter of about 180 fs (rms). (c) and (d) are the slow drifts of beam arrival time measured within a time window of 15 minutes. 
    \label{Fig.5}}
    \end{figure}    

It is worth pointing out that the phase jitter in the buncher will lead to timing jitter after bunch compression \cite{RFC1, RFC3, SJTUstreaking}. Under full compression condition, 50 consecutive measurements of the beam profile are shown in Fig.~5a, and the arrival time jitter determined by recording the fluctuations of the beam centroid is estimated to be about 180 fs (rms). From Fig.~5b one can see that the range of the arrival time jitter is larger than the useful window of linear streaking. With circular deflection, however, all the data points can be recorded and accurately mapped to time. The accuracy of the arrival time measurement is mainly affected by the fluctuation of the centroid divergence of the electron beam, resulting in temporal offset in the measurement. In this experiment, the shot to shot fluctuation of the beam centroid divergence is measured to be about 13 $\mu$rad, corresponding to an uncertainty of 3 fs in beam arrival time determination. It is worth mentioning that in this measurement the dielectric tube has a large aperture that allows all the electrons to pass through. Together with the large dynamic range, this THz oscilloscope is well suited as an on-line timing tool to measure and correct the timing jitter to improve the temporal resolution in pump-probe experiments, in particular rf buncher based keV UED \cite{RFC3, SiwickC, MartinC} and MeV UED \cite{UED3, UCLA, THU, OSAKA, SJTU, BNL, SLAC, DESY}.

The large dynamic range of this THz oscilloscope also allows us to unambiguously record the slow drift of the timing over a much longer time. In this measurement the jitter is integrated over 1000 pulses and is recorded at a step of 20 seconds. In contrast to the short term jitter where the arrival time fluctuates from shot to shot (Fig.~5a), the long term jitter shows a slow drift and oscillation behavior (Fig.~5c). In this measurement, the arrival time drifted by about 0.5 ps during the measurement of 15 minutes. The measured slow timing drift may be incorporated in a feedback system to maintain the long term timing stability, which can be crucial for certain experiments (see, e.g. \cite{SLACgasphaseCF3I} that require long data acquisition time.   

In conclusion, we have demonstrated a novel and practically useful method for recording time information of relativistic electron beams. In our experiment we have achieved 24 fs resolution in beam temporal profile measurement and 3 fs accuracy in determining the beam arrival time with a THz pulse having field strength of about 150 kV/cm and a dielectric tube with $D$=0.87 mm. Because the impedance of the deflection mode scales as $D^{-3}$ and THz pulse with field exceeding 1 MV/cm has been achieved \cite{GV}), the resolution of THz oscilloscope can be potentially extended to sub-femtosecond with stronger THz pulse and smaller aperture dielectric tube. It should be noted that the temporal resolution in this measurement scales inversely with the square root of electron energy, so it is also straightforward to apply this THz oscilloscope for characterizing electron beams with higher energy. Furthermore, this THz oscilloscope may be directly used to enhance the temporal resolution of UED. For instance, the diffracted beam just downstream of the sample may be swept with the THz oscilloscope, forming ring patterns for operating the UED in movie mode \cite{Mourou1982, UCLAmovie, THUmovie, keVmovie}. In this configuration, the temporal resolution in UED is no longer limited by electron bunch length and timing jitter, but rather by the temporal resolution of the THz oscilloscope, allowing one to reach femtosecond temporal resolution with picoseond electron beam. We expect this THz oscilloscope to have wide applications in many areas of researches. 

This work was supported by the Major State Basic Research Development Program of China (Grants No. 2015CB859700) and by the National Natural Science Foundation of China (Grants No. 11327902, 11504232 and 11721091). One of the authors (DX) would like to thank the support of grant from the office of Science and Technology, Shanghai Municipal Government (No. 16DZ2260200 and 18JC1410700).\\
* dxiang@sjtu.edu.cn \\
* jzhang1@sjtu.edu.cn

\pagebreak

\end{document}